# Experimental Study of A Novel Variant of Fiduccia Mattheyses(FM) Partitioning Algorithm

[1]Mitali Sinha, [2]Rakesh Mohanty, [4]Prachi Tripathy
Department of Computer Science and Engineering
Veer Surendra Sai University of Technology, Burla,
Sambalpur, Odisha, India-768018
[1]mitalisinha.148@gmail.com, [2]rakesh.iitmphd@gmail.com,
[4]prachitripathy@gmail.com

[3]Suchismita Pattanaik
Department of Electronics and Communication Engineering
Sambalpur University Institute of Information Technology,
Jyoti Vihar, Burla, Sambalpur, Odisha, India-768019
[4]spattanaik.vlsi@gmail.com

*Abstract*— Partitioning is a well studied research problem in the area of VLSI physical design automation. In this problem, input is an integrated circuit and output is a set of almost equal disjoint blocks. The main objective of partitioning is to assign the components of circuit to blocks in order to minimize the numbers of inter-block connections. A partitioning algorithm using hypergraph was proposed by Fiduccia and Mattheyses with linear time complexity which has been popularly known as FM algorithm. Most of the hypergraph based partitioning algorithms proposed in the literature are variants of FM algorithm. In this paper, we have proposed a novel variant of FM algorithm by using pair wise swapping technique. We have performed a comparative experimental study of FM algorithm and our proposed algorithm using two datasets such as ISPD98 and ISPD99. Experimental results show that performance of our proposed algorithm is better than the FM algorithm using the above datasets.

## I. INTRODUCTION

Hypergraph partitioning is a NP-hard problem[4]. Though hypergraph partitioning has extensive applications in various fields such as data-mining, job scheduling, image processing, improving page fault and VLSI design, a number of heuristic algorithms were developed with polynomial time-complexity. Fiduccia and Mattheyses (FM) algorithm [5] is a basic hypergraph partitioning algorithm with single shift in which time-complexity is linear in nature. In this paper, we studied the FM algorithm and explored its limitations. We proposed a variant of FM algorithm and conducted experimental studies of our proposed algorithm by considering two standard data sets. Through our experiments, we have done comparative performance analysis of our proposed algorithm with the FM algorithm.

*Very Large Scale Integration (VLSI)* is a technique of manufacturing an Integrated Circuit(IC) by integrating thousands of connected electronic components into a single chip. The components may be transistors, resistors, capacitors and inductors etc. A group of connected components can be represented as a *block*. In circuit layout, the length of connections between the components of two different blocks is more than that of the length of connections between the components within the same block. Therefore we have to minimize the number of connections between the components of two different blocks to reduce the cost of wire length.

*A. Hypergraph Partitioning*

Circuit in the form of a graph or hypergraph is provided as the input to a partitioning algorithm. A *hypergraph* is a generalization of graph in which an edge connects any number of vertices and this edge is called as *hyperedge*. Mathematically hypergraph can be represented as *H(V, E)*, where *V* is the set of vertices and *E* is the set of hyperedges. A circuit can be converted to a hypergraph in which a vertex of hypergraph represents component of the circuit and a hyperedge represents the set of components which share the same signal known as *net*. A set of nets which represent a circuit is known as *netlist*. Two vertices of a hypergraph are said to be *neighbor* if both belong to at least one common net. A circuit and its netlist representation are shown in fig. 1(a) and fig. 1(b) respectively.

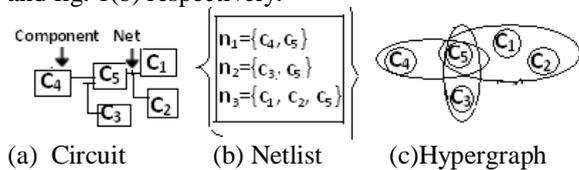

(a) Circuit      (b) Netlist      (c)Hypergraph

Fig.1 Input representation of a partitioning algorithm

The netlist in fig. 1(b) contains three nets $n_1$, $n_2$, and $n_3$. $n_1$ contains components $c_4$, $c_5$ and output of $c_4$ is provided as input to $c_5$. Similarly $n_2$ contains $c_3$, $c_5$ and $n_3$ contains $c_1$, $c_2$ and $c_5$. The Netlist is represented as a hypergraph $H(V, E)$ as shown in fig.1(c) in which V={$c_1$, $c_2$, $c_3$, $c_4$, $c_5$} and E={$n_1$, $n_2$, $n_3$}.

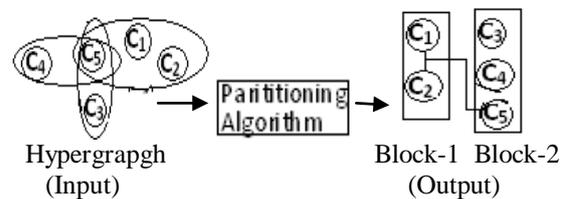

Hypergrapgh           Block-1 Block-2
(Input)                     (Output)

Fig. 2 Input and Output of a Hypergraph partitioning

We consider a hypergraph as shown in Fig.2 as the input to a partitioning algorithm and divide it into two approximately equal blocks. Here the components $c_4$, $c_5$, $c_3$ are present in Block-1 and $c_1$, $c_2$ are present in Block-2. The number of netcuts of this partitioning is one because the components of net $n_3$ are present in both the blocks.

*B. Literature Review*

In order to solve the partitioning problem in VLSI context, the first graph bi-partitioning algorithm was proposed by Kernighan and Lin[1], popularly known as KL algorithm. The time-complexity of KL algorithm is $O(n^3)$ where $n$ is the number of vertices of the input graph. A faster KL algorithm was introduced in [6]. As reported in [7], graph is not a proper representation of a circuit because it cannot correctly convert a net to an edge or a set of edges. The most correct representation of a circuit is hypergraph. A hypergraph partitioning algorithm was proposed by Fiduccia and Mattheyses [5] in the year 1982. The main advantage of this algorithm is its linear time-complexity with respect to the size of the circuit.

A number of variants of FM algorithm with improved performance were developed in [8]-[9]-[10]-[11]. Alpert and Kahng have done a comprehensive survey on netlist partitioning in [3]. A new class of partitioning algorithms known as *2-phase FM*, has been mentioned in [12]-[13]-[14]-[15]. FM algorithm has been extended to various *multi-level FM* algorithms [16]-[17]-[18]-[19]-[20] for better result in terms of solution quality and run time.

*C. Our Contribution*

In this paper we have proposed a novel variant of FM algorithm by using the idea of pair wise swapping of vertices in hypergraph partitioning. Initially a hypergraph is partitioned in to two blocks of roughly equal size by randomly assigning the vertices of hypergraph to each of the blocks. Then vertices are selected in pair wise manner and swapped in order to reduce the total number of netcuts. We have developed a formula for reduction in netcuts due to pair-wise swapping of components in hypergraph partitioning. We have made a comparative performance analysis of our proposed variant of FM algorithm with FM algorithm using two data sets such as ISPD98 and ISPD99 benchmark circuits. Our experimental results show that our proposed algorithm outperforms FM algorithm.

*D. Organization of Paper*

FM algorithm and its pseudo-code are presented in section II. Section III contains our proposed variant of FM algorithm and its pseudo-code. Our experimental study and results are shown in section IV. Section V presents the conclusion and future work.

## II. FM ALGORITHM

In this section we introduce some basic notations and definitions then we represent the pseudo-code of FM algorithm.

*A. Notations and Definitions*

Let $N_{cut}$ be the total number of nets which are cut. *Cutset* be the set of nets which are cut and $n(c_k)$ be the set of nets connected to $c_k$. $P$ be the maximum number of nets to which any component is connected.

*Block Size($S(B_i)$):* The number of components present in a block is defined as the block size.

*Complementary block:* If partitioning of a netlist contains two disjoint blocks $B_1$, $B_2$ and a component $c_k$ is present in $B_1$, then $B_2$ is called the complementary block of $c_k$.

*Unlocked Component:* When a component is free to move from its current to its complementary block, it is called free or unlocked component.

*Locked Component:* When a component is shifted from its current block to its complementary block, it will not be considered for further movement. So it is locked.

*Gain value ($G(c_k)$):* The gain value of a component $c_k$ is the number of reductions in nets from Cutset if it is moved from its current block to its complementary block. it is calculated as follow.

Let $N_{c_k}$: Number of net which have only one component i.e $c_k$ in the current block of $c_k$.

Let $N'_{c_k}$: Number of nets which contain component $c_k$ and completely present in the current block of $c_k$.

$$G(c_k) = N_{c_k} - N'_{c_k}$$

*Gain bucket*: Gain bucket is used to sort the gain values of the components present in a block. Its index ranges from $-P$ to $+P$. The $K^{th}$ index of gain bucket contains a linked list of components having gain value K.

*Update_neighbor's Gain of($c_k$):* This function update the gain values of all unlocked components which are neighbors of $c_k$ [5] and this update will be reflected in the gain buckets.

*Make_Unlock($c_k$):* This function is used to unlock a component $c_k$.

*Make_Lock($c_k$):* This function is used to lock a component $c_k$ and delete $c_k$ from its gain bucket.

*B. Pseudo Code of FM Algorithm*

The first hypergraph bi-partitioning algorithm is the FM algorithm [5] with linear time complexity. It starts with a random initial partitioning of the hypergraph H into two almost equal size blocks $B_1$ and $B_2$ and $N_{cut}$ is calculated. At the beginning of the process, all the components are made unlocked and the gain value of each component is calculated. Components of each block are sorted using bucket sorting according to their gain values in order avoid unnecessary search for the component having maximum gain value.

A component $c_k$ with highest gain value is selected to move from its current block to its complimentary block and remains locked throughout the process. The size of $c_k$'s current block should be greater than or equal to its complimentary block. After $c_k$ is moved, the gain values of its all unlocked neighbors are updated in their respective gain bucket for next move and $N_{cut}$ is recorded at that point. This is continued until all components are locked.

This entire process is called a pass. When a component is locked, it cannot be considered for further move within that pass. At the end of a pass, the point at which the optimal $N_{cut}$ was achieved is selected and the moves of all components after that point are cancelled. The partitioning result of one pass is given as input to next pass. This process is continued till improvement in $N_{cut}$. Finally the optimal $N_{cut}$ is achieved.

After a comprehensive study and analysis of FM algorithm the following limitations are observed.

*C. Limitations of FM Algorithm*

When more than one component has same gain value then FM algorithm randomly choose any one component for shifting. So it does not always provide optimal result. Component's move operation is highly influenced by the balancing constraint of block [21]. FM algorithm uses the technique of single shifting of component instead of Pair wise swapping but pair-wise swapping provides better result than single shifting of component [6]. First limitation is addressed by many other proposed algorithms described in [8]-[10]-[11]. In our work, we have addressed the last two limitations by developing a novel variant of FM algorithm.

### III. PROPOSED VARIANT OF FM ALGORITHM

An algorithm that swaps node pairs can provide a better $N_{cut}$ improvement than one that shifts a single node at a time [6]. In this paper we have applied pair wise swapping of components on hypergraph partitioning by proposing a novel variant of FM algorithm. In this variant of FM algorithm, two components from each block are swapped so that this pair produces the maximum reductions in nets from Cutset than any other pair as proposed in [6] for graph partitioning. Before presenting our proposed variant of FM algorithm we introduce some definitions and notations as follows.

*Critical net($n_c$)*: If any component of a net is shifted from its current block to its complementary block and as a result the net is being removed from Cutset then such a net is called *critical net*. In fig.4 $n_3=\{c_1, c_2, c_5\}$ is a critical net because $n_3$ is being removed from Cutset due to shifting of $c_5$ from $B_1$ to $B_2$.

```
Input   : H(V,E) → A circuit as hypergraph
Output  : B_1, B_2 → A partition with optimum N_cut
Notations:
  u_i     → {u_i: u_i ∈ B_1 for i = 1 to S(B_1)}
  v_j     → {v_j: v_j ∈ B_2 for j = 1 to S(B_2)}
  GB_i    → Gain bucket of block B_i.
  G_total → Gain due to shifting of u_i and v_j.
  G_max(B_i) → maximum gain value of any c_k such that
               c_k ∈ GB_i for i=1, 2.
  U_Bi    → set of unlocked components in B_i
  L_Bi    → set of locked components in B_i.
FM pass :
  Step 1:
  Do a Random Initial partitioning (H(V, E)) into blocks
  B_1, B_2 such that B_1∩B_2=∅, B_1∪B_2=V, |S(B_1)-S(B_2)| ≤ 1
  Step 2:
  Make_Unlock(c_k); ∀c_k ∈ B_1∪B_2
  U_B1={u_i | ∀u_i ∈ B_1} ; U_B2={v_j | ∀v_j ∈ B_2}; L_B1=∅; L_B2=∅;
  Step 3:
  Calculate G(c_k); ∀c_k ∈ B_1∪B_2 ;
  Sort the components of each block B_i according to
  their gain values in their respective gain bucket GB_i.
  Step 4:
```

```
G_total=0 ; N_opt= N_cut ; G_max(B_i)=maximum{G(c_k)|∀c_k ∈ GB_i}
Until (V− (L_B1 ∪ L_B2)=∅)
  Begin
    If((G_max(B_1)≥G_max(B_2)) and (S(B_1)≥S(B_2))
      Randomly select a u_i from G_max(B_1) index of GB_1
      G_total=G_total + G(u_i); N_opt= N_opt − G(u_i);
      Make_Lock(u_i) ; U_B1=U_B1 −u_i; L_B2 =L_B2 ∪ u_i ;
      Update_neighbor's Gain of(u_i);
      G_max(B_1)=maximum{G(u_i)|∀u_i ∈ GB_1}
    Else
      If ((S(B_2)≥S(B_1))
        Randomly select a v_j from G_max(B_2) index of GB_2
        G_total=G_total + G(v_j); N_opt= N_opt − G(v_j);
        Make_Lock(v_j) ; U_B2=U_B2 − v_j; L_B1 =L_B1 ∪ v_j;
        Update_neighbor's Gain of(v_j);
        G_max(B_2)=maximum{G(v_j)|∀v_j ∈ GB_2}
      Else
        Randomly select a u_i from G_max(B_1) index of GB_1
        G_total=G_total + G(u_i); N_opt= N_opt − G(u_i);
        Make_Lock(u_i) ; U_B1=U_B1 −u_i; L_B2 =L_B2 ∪ u_i ;
        Update_neighbor's Gain of(u_i);
        G_max(B_1)=maximum{G(u_i)|∀u_i ∈ GB_1}
End
Step 5:
Select the best seen configuration where G_total is maximum
or N_opt is minimum
if( G_total > 0 ) then Shift the components to
their original position after reaching the best configuration.
```

Fig.3 Pseudo-code of FM algorithm

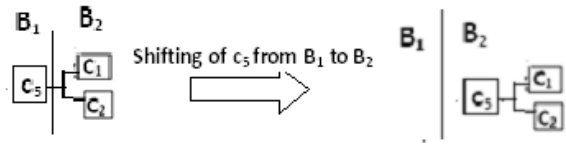

Fig. 4 Critical net

*Correct_term:* Correct_term is a non-negative integer value which represents the number of common nets of $u_i$, $v_j$ in Cutset both before and after swapping of $u_i$, $v_j$. Pseudo-code for correct term is shown in Fig. 5.

```
Correct_term Algorithm
  Begin
    Correct_term=0;
    For each n_c
      If( u_i ∈ n_c and v_j ∈ n_c)
        If (cardinality of n_c == 2)
          Correct_term=Correct_term+2;
        Else
          Correct_term=Correct_term+1;
  End
```

Fig.5 Pseudo-code for correct term

*Gain due to pair-wise swapping ($G(u_i, v_j)$):*

$$G(u_i, v_j)=G(u_i) + G(v_j) - correct\_term.$$

*A. Variant of FM Algorithm*

We present pseudo-code for our proposed variant of FM algorithm as shown in Fig.6. Two blocks can be visualized as an $m*m$ matrix $M$ with one axis of $M$ is the sorted components of $B_1$ and other axis is the sorted components of $B_2$ according to their gain values. Each element of *M* corresponds to component pair $(u_i, v_j)$. In this variant, $u_i$ and $v_j$ from each block are selected so that $(u_i, v_j)$ pair provides highest gain value than any other pair. The pseudo-code of FM_Variant for best pair selection in hypergraph partitioning is described in Fig.7 as mentioned in [6] for graph. The worst case time complexity for finding the first non-neighbor component of $u_1$ is $O(d)$ i.e $b=d$ where $d$ is the maximum degree of any component if the hypergraph is visualized as a clique based graph[3]. In the worst case, the gain values of all the neighbor components of $u_1$ will be greater than gain value of non-neighbor components of $u_1$. So $b$ value will be repeated $d$ times to reach the 1st non-neighbor component in the 1st while loop condition in the Fig.7. After this, a for loop is continued at most $(d–1)$ times and within the for loop a while loop is continued at most $d$ times. The total time complexity is $(d+(d–1)*d)$. The worst case time complexity to get best pair is $O(d^2)$. Hence selecting $m$ best pairs time complexity is $O(d^2*m) \approx O(e*d)$. $d <<< m$.

```
Notations:
 k: A integer value
 Dv: A Dummy vertex
 GPS: gain value due to pair-wise swapping of components
 m=⌈S(V)/2⌉
FM_Variant Pass:
Step 1:
 If S(V)=2*k+1 then V'=V U Dv and e(∀v, Dv)∈H(E), v∈V
Step 2:
 Do a Random Initial partitioning (H(V, E)) into blocks B1
 and B2 such that B1∩B2=∅, B1UB2=V and S(B1)=S(B2)=m
Step 3:
 Make_Unlock(ck); ∀ck ∈B1UB2
 UB1={ui | ∀ui ∈B1}; UB2={vj | ∀vj ∈B2}; LB1=∅; LB2=∅;
Step 4:
 Calculate G(ck) for ∀ck ∈B1UB2
 Sort the components of each block Bi according to
 their gain value such that G(u1)≥G(u2)≥…≥G(um) in GB1
 and G(v1)≥G(v2)≥…≥G(vm) in GB2.
 GPS=0; Nopt =Ncut;
  Until((V— (LB1 U LB2))=∅)
   Begin
    Select ui ∈ GB1 and vj ∈ GB2 using a pair-wise swapping
    technique.
     GPS = GPS + G(ui, vj); Nopt = Nopt — G(ui, vj);
     Make_Lock(ui); UB1=UB1 — ui; LB2 =LB2 U ui;
     Update_neighbor's Gain of(ui);
     Make_Lock(vj); UB2=UB2 — vj;LB1 =LB1 U vj;
     Update_neighbor's Gain of(vj);
   End
Step 5:
 Select the best configuration where GPS is maximum or
 Nopt is minimum
 If(GPS > 0) then Shift the components to their original
 position after reaching the best configuration.
```

Fig.6 Pseudo-code of FM_Variant

```
G(u1)≥G(u2)≥…≥G(um) in GB1
G(v1)≥G(v2)≥…≥G(vm) in GB2
Gbestpair= Gain value due to best pair
 b=1;
 while(vb is neighbor of u1) and (K<m))
  b++;
 Gbestpair =G(u1, vb);
 (u, v)=( u1, vb) //now best pair (u1, vb)
 For j=1 to (b-1)
  i=0
  do
  {
   i++
   if(G(ui, vj)> Gbestpair)
    Gbestpair = G(ui, vj);
    (u, v)=(ui, vj); // now best pair (ui, vj)
   If(j=1 and G(ui)+G(vj)<Gbestpair
    return (u, v);
  }
  while(vj is neighbor of ui and i<=m);
  return (u, v);// best pair is returned
```

Fig.7 Pseudo code for best pair selection for FM_variant

## IV. EXPERIMENTAL STUDY

In our experimental study, we have evaluated the performance of FM algorithm and our proposed FM_variant algorithm by computing optimal $N_{cut}$. The above two algorithms are tested using two large datasets called as ISPD98 and ISPD99 benchmark circuits. In our experiments we randomly select a component when more than one component has the same gain value.

### A. Experimental Setup

The source code for the implementation is developed in 'C++' language and windows operating system environment. The compiler is 32-bit compiler (Dev C++ Version 4.9.9.2). RAM size is 2GB and processor speed is 2 GHz. Input to the program is an IBM file. Then the components and nets are extracted from the file using our program. The size of all components same in input file. The output of the program is optimal $N_{cut}$ after partitioning the components equally between the two blocks.

### B. Input Dataset

The ISPD98 circuit benchmark is the largest dataset which is maintained by the Collaborative Benchmarking Laboratory. The ISPD98 circuit benchmark contains 18 types of files. These are IBM01 to IBM18. Each file comes with three formats such as .net, .are and .netD. Another version of ISPD benchmark circuit is ISPD99. This benchmark circuit contains 9 types of file. Each file having 4 version with .netD or .are format. These dataset are freely in the website
http://vlsicad.ucsd.edu/UCLAWeb.html

### C. Experiments Performed

We have performed the two experiments using two different datasets ISPD98 and ISPD99. For the experiments we have defined *gain(μ)* as follow.

$$gain(\mu) = \{(optimal\ N_{cut}(FM) - optimal\ N_{cut}(FM\ variant)) / optimal\ N_{cut}(FM)\} * 100.$$

The Performance of FM_variant is observed to be better if the gain is higher and positive. In the first experiment we have computed the optimal $N_{cut}$ of FM and our proposed variant of FM algorithm by considering ISPD98 as input dataset and compared the optimal $N_{cut}$ of both the algorithms. In the second experiment, we have computed optimal $N_{cut}$ of FM and our proposed variant of FM algorithm by taking ISPD99 as input dataset and compared the optimal $N_{cut}$ of both the algorithms.

EXPERIMENT-1: ISPD98 AS INPUT DATASET

In this Experiment we have considered eighteen different files of ISPD98 benchmark circuit. We have computed the optimal

$N_{cut}$ of FM and our proposed variant of FM algorithm as shown in Table I.

EXPERIMENT-2: ISPD99 AS INPUT DATASET
In this experiment we have taken nine different files of ISPD99 benchmark circuit and each file having 4 different versions. We have computed the optimal $N_{cut}$ of FM and our proposed variant of FM algorithm as shown in Table II.

For experiment-1 and experiment-2 we plot the graph by considering file's name of dataset in the X-axis and *gain(μ)* in Y-axis as shown in fig.8 and fig.9 for ISPD98, ISPD99 respectively.

TABLE I. FM VS FM VARIANT OF ISPD98

| File 'name | Initial $N_{cut}$ | optimal $N_{cut}$(FM) | optimal $N_{cut}$ (FM_ variant) | Gain (μ) |
|---|---|---|---|---|
| IBM01 | 9151 | 1534 | 858 | 44.06 |
| IBM02 | 13443 | 1595 | 529 | 66.8 |
| IBM03 | 17422 | 4013 | 2885 | 28.1 |
| IBM04 | 20643 | 4327 | 1016 | 76.5 |
| IBM05 | 18895 | 6881 | 3402 | 50.5 |
| IBM06 | 22798 | 5721 | 1475 | 74.2 |
| IBM07 | 32044 | 7028 | 2516 | 64.2 |
| IBM08 | 33499 | 9242 | 3321 | 64.06 |
| IBM09 | 40173 | 10438 | 2809 | 73.08 |
| IBM10 | 50647 | 10413 | 2454 | 76.43 |
| IBM11 | 54221 | 12893 | 4086 | 68.3 |
| IBM12 | 52102 | 14508 | 4312 | 70.27 |
| IBM13 | 23076 | 5275 | 1159 | 78 |
| IBM14 | 101990 | 22990 | 11257 | 51.04 |
| IBM15 | 125878 | 29037 | 15149 | 47.82 |
| IBM16 | 129985 | 37057 | 7268 | 80.4 |
| IBM17 | 131364 | 42226 | 10062 | 76.2 |
| IBM18 | 139169 | 36949 | 3055 | 91.73 |

TABLE II. FM VS FM VARIANT OF ISPD99

| File's Name | Initial $N_{cut}$ | optimal $N_{cut}$(FM) | optimal $N_{cut}$(FM variant$_2$) | Gain (μ) |
|---|---|---|---|---|
| IBM01A | 9213 | 2148 | 364 | 83.1 |
| IBM01B | 4958 | 685 | 124 | 82 |
| IBM01C | 4878 | 800 | 495 | 38.13 |
| IBM01D | 4985 | 1268 | 295 | 76.74 |
| IBM06A | 22889 | 5147 | 1328 | 74.2 |
| IBM06B | 9962 | 2565 | 948 | 63.41 |
| IBM06C | 14558 | 3044 | 963 | 68.4 |
| IBM06D | 8693 | 2360 | 829 | 65 |
| IBM09A | 40187 | 9966 | 3026 | 69.63 |
| IBM09B | 33104 | 8064 | 3830 | 52.51 |
| IBM09C | 36086 | 8022 | 4706 | 41.34 |
| IBM09D | 33809 | 8343 | 2674 | 67.95 |
| IBM10A | 50751 | 11989 | 2865 | 76.1 |
| IBM10B | 20021 | 4259 | 2551 | 40.1 |
| IBM10C | 32876 | 7819 | 2322 | 70.3 |
| IBM10D | 21601 | 4938 | 1394 | 71.77 |
| IBM11A | 54079 | 12833 | 4408 | 65.65 |
| IBM11B | 27379 | 5701 | 3006 | 47.27 |
| IBM11C | 29364 | 7613 | 2744 | 63.95 |
| IBM11D | 24600 | 5471 | 3219 | 41.16 |
| IBM12A | 51921 | 13339 | 3671 | 72.48 |
| IBM12B | 29498 | 6981 | 2755 | 60.53 |
| IBM12C | 25109 | 5533 | 2018 | 63.53 |
| IBM12D | 22915 | 6805 | 1841 | 72.94 |
| IBM13A | 66251 | 15645 | 3197 | 79.56 |
| IBM13B | 32990 | 6813 | 499 | 92.67 |
| IBM13C | 35728 | 4461 | 1427 | 68 |
| IBM13D | 31019 | 6766 | 2019 | 70.16 |
| IBM16A | 129950 | 36869 | 6844 | 81.44 |
| IBM16B | 72220 | 19083 | 2653 | 86.1 |
| IBM16C | 61793 | 18269 | 3047 | 83.32 |
| IBM16D | 48717 | 14341 | 3084 | 78.5 |
| IBM17A | 131753 | 43544 | 9609 | 77.93 |
| IBM17B | 74174 | 24782 | 2186 | 91.17 |
| IBM17C | 62567 | 18965 | 4157 | 78.1 |
| IBM17D | 41472 | 11707 | 3802 | 67.52 |

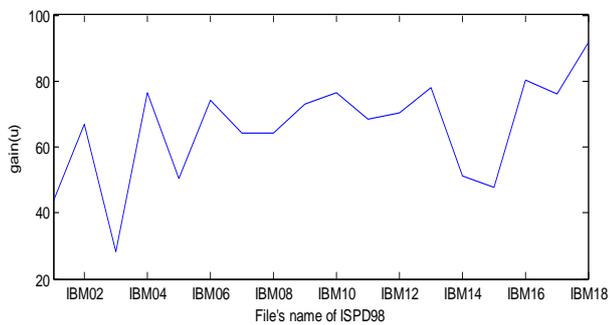

Fig.8 FM VS FM_VARIANT OF ISPD98

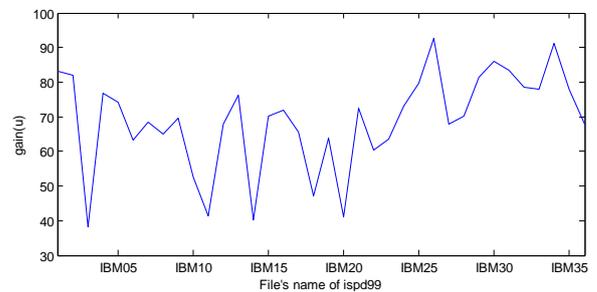

Fig.9 FM VS FM_VARIANT OF ISPD99

## V. CONCLUSION AND FUTURE WORK

In this work we have proposed the variant of FM algorithm using a Pair-wise Swapping technique. We have conducted an experimental study to evaluate the performance of our proposed algorithm and FM algorithm by considering two input datasets such as ISPD98 and ISPD99 benchmark circuits. From experimental result, we observed that our proposed algorithm outperforms FM algorithm.

In future work, we can consider and apply FILO technique for selections of components from gain bucket in our proposed algorithm and compare its performance with FM-LIFO [13]. As reported in [13], FILO technique provides better result than random and FIFO technique. Our proposed variant of FM algorithm can also be enhanced by using multi-level technique.